\documentclass[12pt, epsf]{article}
\usepackage{amsfonts,amssymb,amsmath,amscd, graphicx}

\title{Simple solutions to the Einstein Equations in spaces with unusual topology}
\author{Mihai Bondarescu
\footnote{California Institute of Technology MC 452-48, Pasadena, CA 91125, U.S.A., email 
mihai@theory.caltech.edu, web page http://www.theory.caltech.edu/\~{}mihai}}

\begin{document}

\begin{titlepage}

\maketitle

\begin{abstract}

We discuss simple vacuum solutions to the Einstein Equations in five dimensional space-times 
compactified in two different ways. In such spaces, one black hole phase and more then one black string 
phase may exist. Several 
old metrics are adapted to new background topologies to yield new solutions to the Einstein Equations. 
We then briefly talk about the angular momentum they may carry, the horizon topology and phase 
transitions that may occur.

\end{abstract}

\end{titlepage}

\section{Introduction}

Higher dimensional gravity has been a very active area of theoretical physics lately. Long ago, the 
Schwarzschild solution has been generalized to higher dimensions by Tangherlini 
\cite{Tan}. Higher dimensional metrics that satisfy the Einstein Equations can be obtained from lower 
dimensional ones by tensoring a Ricci-Flat space to a known vacuum solution. In this fashion,
one can build things like black strings or black branes by tensoring some flat euclidean space to a 
black hole solution. A black string living in 5 dimensions would then have the metric: 

\begin{equation}
ds^2_{BS}=ds^2_{BH}+dx_4^2  
\label{eq1}
\end{equation}
where $ds^2_{BH}$ is any black hole metric. 

The uniform black strings have been shown to be unstable when not thick enough \cite{GL, Barak} and 
stationary 
non-uniform solutions have been found \cite{Kudoh1, Kudoh2}. Numerical evolutions of black strings have 
also shown them to be unstable and suggest that thin black strings decay into black holes 
\cite{Choptuik}. There is still 
some controversy over the end point of the black string instability \cite{Horowitz} as in classical 
General Relativity the horizon topology cannot easily change. 

Black ring solutions that have horizon topology $S^2\times S^1$ 
in asymptotically flat five dimensional space-time and violate black hole uniqueness theorem were recently 
discovered \cite{Emparan1, Emparan2, Emparan3, Emparan4, Elvang}. 

So far, solutions of the Einstein Equations like black holes, black strings and branes or gravitational 
waves have only been studied in space-times that are either non-compact or 
periodically compactified. 
In this essay I will take a few baby steps away from the beaten path and 
explore a few simple solutions of Einstein Equations in spaces with antiperiodic compactification and 
spaces where compactification is done by identifying parallel hyperplanes after rotating them by 
some angle $\alpha$. 

We'll explore solutions with simple metrics and distinct topologies and discuss 
phase transitions that are likely to occur between these black objects. 

The spaces discussed here are not very much like the world we live in. In the limit where the length of 
the  compact dimension becomes zero, the five dimensional space-time antiperiodically compactified 
becomes an orbifold. The same is true for the space with twisted compactification.

Throughout this essay, $x_0$ stands for time and $x_4$ is the compact direction.

\section{Antiperiodic compactification}

Antiperiodic compactification of five dimensional space-time is given by the following 
identification: 

\begin{equation}
(x_0, x_1, x_2, x_3, x_4)= (x_0, -x_1, x_2, x_3, x_4+L)
\label{antiper}
\end{equation}
The resulting space is flat but its non-trivial topology has the following 
implications

-- The space is not orientable. Therefore one cannot have a globally-defined 
volume form. 

-- The hyperplane $x_1=0$, although non-singular is singled out as "special" as 
it is invariant under the identification (\ref{antiper}). 

-- There are no globally defined continuous vector fields with nowhere vanishing $x_1$ component.

-- The $x_1$ component of the momentum of a particle moving in this space is not conserved. 

\begin{figure}

\includegraphics[width=0.75\textwidth]{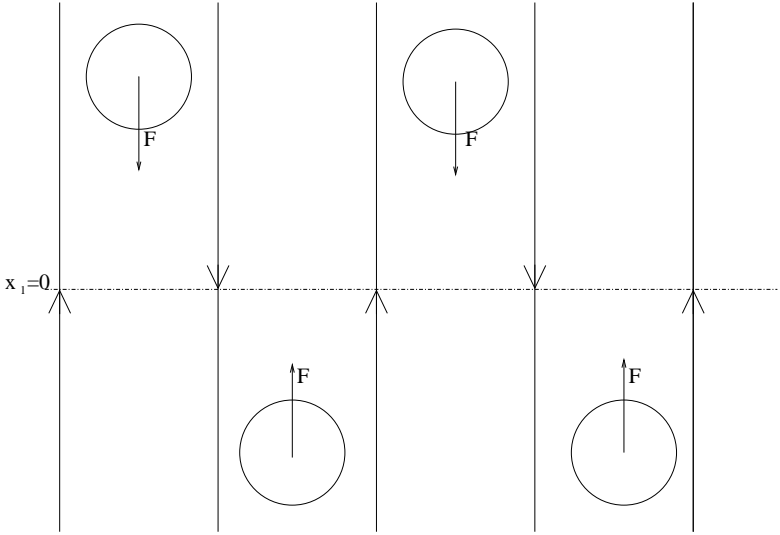}
\caption{An observer near a black hole in an antiperiodically compactified space will see an array of
infinitely many black hole images arranged like above. This solution is neither stationary, nor exact 
as the holes feel an attractive force $F$ toward the $x_1=0$ hyperplane.}
\label{fig:bh}
\end{figure}

Just like in the case of periodic compactification \cite{Barak}, exact solutions for five 
dimensional black holes and black rings cannot be written with our present knowledge. One 
non-rotating, uncharged black hole in this background will be a static solution 
only if its center lies in the $x_1=0$ hyperplane. In this case, the metric is 
actually identical to the one of a black hole sitting in periodically compactified 
space because of its reflection symmetry. 

If it wanders away, the black hole will be attracted toward $x_1=0$ by 
its image and it will undergo an oscillatory motion. (Fig. \ref{fig:bh})

The antiperiodic compactification allows two types of black strings to exist. We'll 
call them Type A and Type B. 

Uncharged uniform non-rotating black string solutions will be stationary only 
if they lie on the $x_1=0$ hyperplane. In this case (Type A, Fig. 
\ref{fig:typeAB}), 
the metric is identical to that of the uniform black string living in 
periodically compactified space:

\begin{figure}

\includegraphics[width=0.49\textwidth]{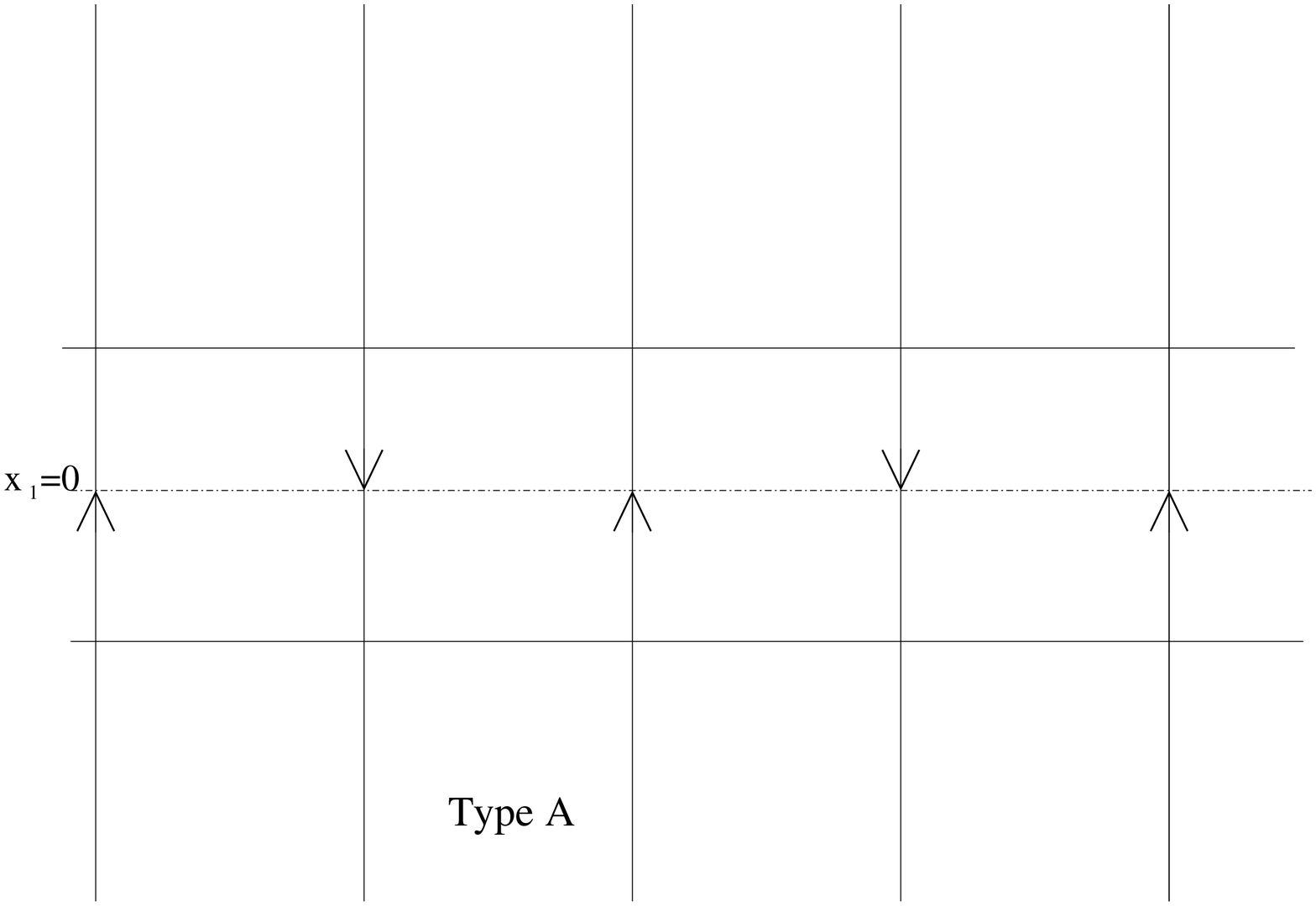}
\includegraphics[width=0.49\textwidth]{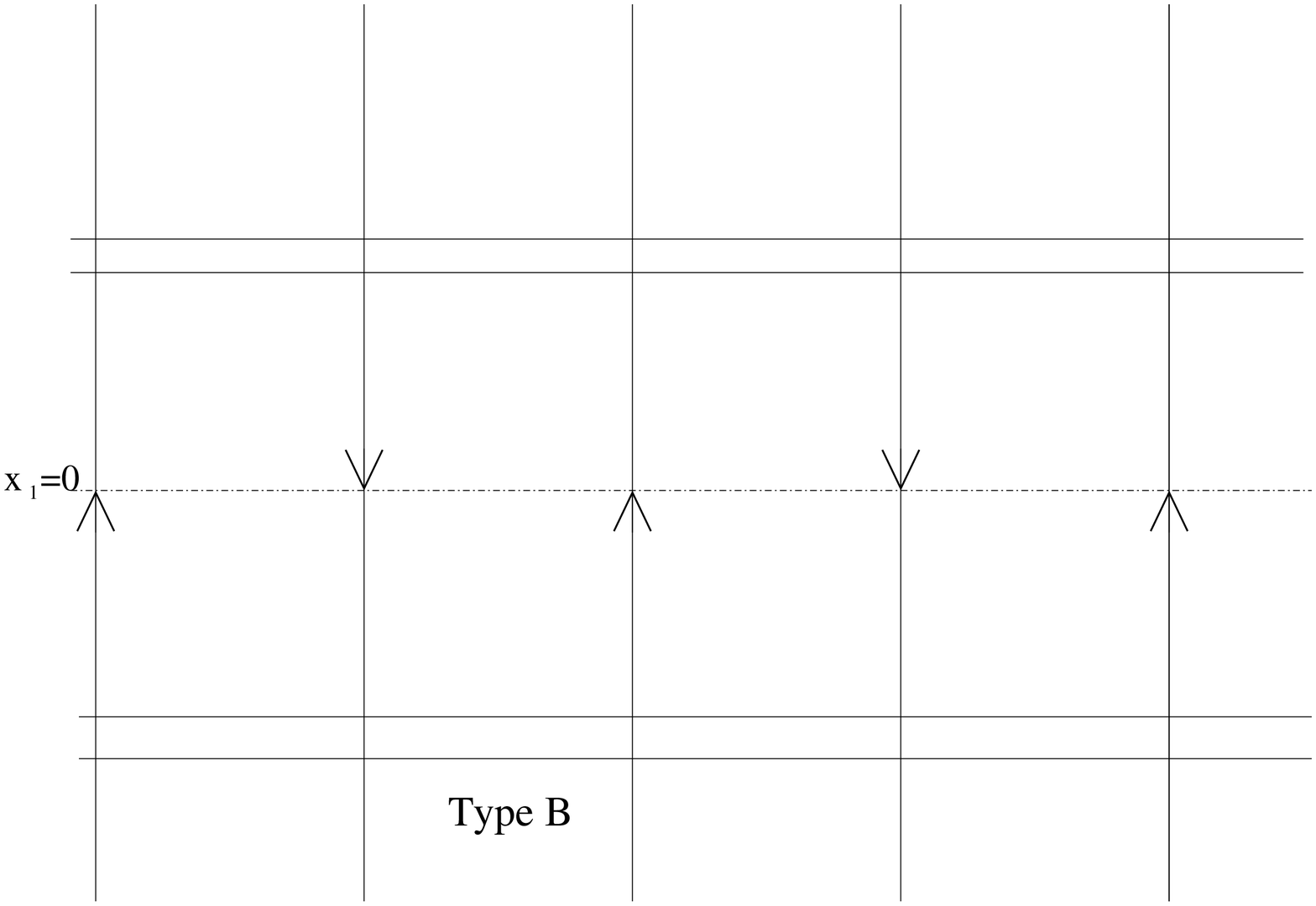}

\caption{Type A and Type B black strings in antiperiodically compactified background.
Picture depicts covering space. The Type A string is both stationary and
exact, but the horizon has topology of a higher dimensional Klein
bottle. Type B is neither stationary nor exact but the horizon has the
"nice" $S^2\times S^1$ topology.}
\label{fig:typeAB}
\end{figure}

\begin{equation}
ds^2 = -\left(1-\frac{r_0}{r}\right)dt^2 + \frac{1}{1-\frac{r_0}{r}}dr^2 + r^2d\theta^2 + 
r^2\sin^2\theta
d\phi^2 + dx_4^2
\label{5Dbs}
\end{equation}
This metric is invariant under (\ref{antiper}). To see this clearly we will 
write (\ref{antiper}) in $(t,r,\theta,\phi,x_4)$ coordinates. The transformation from spherical to 
Cartesian coordinates is 

\begin{equation}
x=r\sin\theta \cos\phi
\end{equation}
\begin{equation}
y=r\sin\theta \sin\phi 
\end{equation}
\begin{equation}
z=r\cos\theta
\end{equation}
Depending on which of $x,$ $y,$ $z$ we pick to be $x_1$, Eq. (\ref{antiper}) becomes one of the 
following:

\begin{equation}
(t, r, \theta, \phi, x_4) = (t, r, \theta, \pi-\phi, x_4+L)
\label{ref1}
\end{equation}

\begin{equation} 
(t, r, \theta, \phi, x_4) = (t, r, \theta, - \phi, x_4+L)
\label{ref2}
\end{equation}

\begin{equation}
(t, r, \theta, \phi, x_4) = (t, r, \pi-\theta, \phi, x_4+L)
\label{ref3}
\end{equation}
The metric (\ref{5Dbs}) is invariant under (\ref{ref1}, \ref{ref2}, \ref{ref3}). Actually, due to 
spherical symmetry, only one of (\ref{ref1}), (\ref{ref2}), (\ref{ref3}) would have been enough but we'll 
need them all a little later when we discuss rotating black strings.  

Unlike its periodically compactified counterpart the horizon topology of this string is 
that of a three dimensional Klein bottle. The surface has no singularities but it 
is non-orientable. 

An infinite rotating black string in uncompactified or periodically compactified space-time will have 
the metric of a Kerr black hole tensored with an extra flat direction: 

\begin{eqnarray}
 ds^2=-\left(1-\frac{2Mr}{\Sigma}\right)dt^2-\frac{4aMr\sin^2\theta}{\Sigma}dt d\phi + 
\frac{\Sigma}{\Delta}dr^2 \Sigma d\theta^2
\label{Kerr} \\
+\left(r^2 + a^2 + \frac{2Mra^2\sin^2\theta}{\Sigma}\right)\sin^2\theta d\phi^2
+dx_4^2
\nonumber
\end{eqnarray}
where 

\begin{equation}
a=\frac{J}{M}
\end{equation}

\begin{equation}
\Delta=r^2-2Mr+a^2
\end{equation}

\begin{equation}
\Sigma=r^2+a^2\cos^2\theta
\end{equation}
One can regard $M$ and $J$ as mass and angular momentum per unit length in the
appropriate units.

The metric (\ref{Kerr}) is only invariant under (\ref{ref3}) and therefore the only uniformly 
rotating 
black strings that can survive in antiperiodically compactified background are those rotating in the $(x_2, x_3)$ 
plane. 

Perturbations that do not alter the horizon topology can move parts of the Type A string 
away from $x_1=0$ hyperplane, but the string horizon will always intersect 
$x_1=0$.

A topologically different black string (Fig. \ref{fig:typeAB}, Type B) is 
obtained by tensoring a flat direction to the space-time containing two 
Schwarzschild black holes at rest with respect to each other and then doing the 
identification (\ref{antiper}). The original metric has the right symmetries to be 
invariant under the identification (\ref{antiper}). The solution has the 
following properties: 

-- The horizon topology is $S^2 \times S^1$, it is orientable and may rotate 
in the planes $(x_1,x_2)$, $(x_1,x_3)$ and $(x_2,x_3)$. The length of the string loop is $2L$, twice 
that of the compact direction. 

-- Although the string may rotate, the total angular momentum in the $(x_1,x_2)$ 
and $(x_2,x_3)$ planes will be zero as an observer living in the antiperiodically compactified 
space will see the two equal sides of the string rotating in opposite directions. 
Only rotation in the $(x_2,x_3)$ plane will yield non-zero total angular momentum. 

-- One cannot write an exact solution for the Type B black string right now because no 
exact solution exists for two black holes colliding head on or at rest with respect to each other in 
4D. 

-- It is not stationary. Black holes will plunge toward each other and coalesce 
in a finite amount of time. One could "fix" this by adding the right cosmological 
constant or some charge. 

\subsection{Phase transitions}

\begin{figure}
\includegraphics[width=0.95\textwidth]{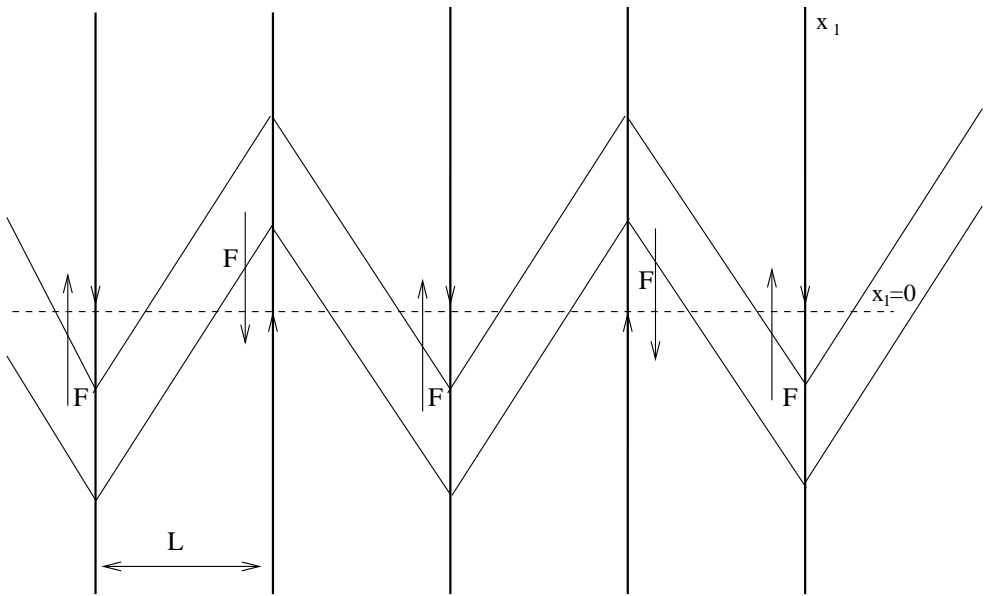}
\caption{A perturbed Type A black string in a world with space compactified as $(x_0,x_1,
x_2, x_3, x_4)=(x_0,-x_1, x_2, x_3, x_4+L)$ may undergo a phase transition
onto either a Type B string or a black hole, both of which can move away easier
from the $x_1=0$ hyperplane.} \label{Fig3}
\end{figure}

When pulled toward $x_1=\pm \infty$ by an external agent the Type A string will deform somewhat 
like in Fig \ref{Fig3}. At some point this string will become more massive then a Type B string of the 
same 
thickness. Then it becomes energetically favorable for the Type A string to decay either in a Type 
B string or - if light enough - to Gregory-Laflamme decay into a black hole. 

Moving the hole or the Type B string away from the $x_1=0$ hyperplane is much easier then moving the Type A 
string away. If we assume the Type A string remains of constant thickness and its mass is proportional 
to its length then for large distances from $x_1=0$ the agent pulling this string away will work against 
a potential 

\begin{equation}
V \sim \sqrt{x_1^2+\left(\frac{L}{2}\right)^2}
\label{vr}
\end{equation}
which, for large enough $x_1$ becomes 

\begin{equation}
V \sim |x_1|
\label{vr2}
\end{equation}
while moving a black hole or a Type B string away from $x_1=0$ when $x_1$ is large enough means only 
working against a potential 

\begin{equation}
V \sim \frac{1}{|x_1|}
\label{v/r}
\end{equation}
which obviously goes to zero for large $x_1$. 

\section{Compactification with a twist}

In previous section we looked at antiperiodically compactified space. 
In this section we'll identify hyperplanes after performing a 
rotation, namely 

\begin{equation}
(x_0, x_1, x_2, x_3, x_4) = 
(x_0, x_1\cos(\alpha)+x_2\sin(\alpha), 
x_2\cos(\alpha)-x_1\sin(\alpha), x_3, x_4+L)
\label{rot}
\end{equation}

These spaces have been studied in \cite{Dowker95} and found to be related to magnetic 
fields in the Kaluza-Klein Theory. 

Although we will deal with arbitrary $\alpha$ here, the particular case of 
(\ref{rot}) when $\alpha=\pi$ is similar to (\ref{antiper})

\begin{equation}
(x_0, x_1, x_2, x_3, x_4) = (x_0, -x_1, -x_2, x_3, x_4+L)
\label{L}
\end{equation}
The transformation (\ref{rot}) leaves points on the line $x_1=x_2=0$ unchanged. 
Therefore, a Schwarzschild or Kerr-Newmann black hole rotating in the $(x_1, x_2)$ 
plane centered on this line will have identical 
metric with its counterpart living in periodically compactified space.

Axisymmetric black strings rotating in the $(x_1, x_2)$ plane or not 
rotating at all will have metrics identical with their 
counterparts living in periodically compactified space (\ref{Kerr}) and respectively (\ref{5Dbs}) just 
because their metrics are invariant under (\ref{rot}) which becomes
\begin{equation}
(t,r,\theta,\phi,x_4)=(t,r,\theta,\phi+\alpha, x_4)
\end{equation} 
If $\alpha = \frac{m}{n}2\pi$, $m, n \in {\mathbb Z}$, $~n \neq 0$  
one can come up with a seemingly odd solution to the Einstein Equations - a set of 
$n$ uniform black strings equally spaced on a large enough circle. Due to the 
compactification (\ref{rot}) the end of one string will be identified with the 
beginning of another and the final solution will have no "free ends". 

Since the space is orientable, one can have the whole solution rotate at the exact 
speed necessary to keep the strings at a constant distance from the origin.  

If one neglects gravitational radiation, this becomes a rather strange 
solution, even stationary in the corotating frame. 

When the string (\ref{eq1}) is taken away from the origin, its length must 
increase. 
Neglecting the perturbations the string causes to the background geometry, the 
minimum length of the string situated a distance $r$ away from the origin is 
$\sqrt{(2r\sin(\frac{\alpha}{2}))^2+L^2}$. 
For large $r$ and $\alpha \neq 2k\pi$, $k \in {\mathbb Z}$ the string length and hence its mass becomes 
proportional to $r$ and it becomes energetically favorable for the string to either Gregory-Laflamme 
decay into a black hole or, if massive enough, a string that winds around the compact direction $n$ 
times. This phase transition will be rather interesting and unusual for large $n$.

\section{Acknowledgments}

I thank Anura Abeyesinghe, Henriette Elvang, Sergei Gukov, Gary Horowitz, Anton Kapustin, Tristan McLoughlin, Andrei 
Mikhailov, Graeme Smith, Iulia Susnea and Kip Thorne for useful discussions, advice, support and 
friendship. This work was supported by a Caltech Teaching Assistantship with David Politzer.


\begin{thebibliography}{99}

\bibitem{Tan}
F.R. Tangherlini,
``Schwarzschild field in n dimensions and the dimensionality of space problem,''
Nuovo Cim. 27(1963) 636

\bibitem{GL}
R.~Gregory and R.~Laflamme,
``Black strings and p-branes are unstable,''
Phys.\ Rev.\ Lett.\  {\bf 70}, 2837 (1993)
[arXiv:hep-th/9301052].

\bibitem{Barak}
B.~Kol,
``The phase transition between caged black holes and black strings: A
review,''
arXiv:hep-th/0411240.

\bibitem{Kudoh1}
H.~Kudoh and T.~Wiseman,
``Connecting black holes and black strings,''
arXiv:hep-th/0409111.

\bibitem{Kudoh2}
H.~Kudoh and T.~Wiseman,
``Properties of Kaluza-Klein black holes,''
Prog.\ Theor.\ Phys.\  {\bf 111}, 475 (2004)
[arXiv:hep-th/0310104].

\bibitem{Choptuik}
  M.~W.~Choptuik, L.~Lehner, I.~Olabarrieta, R.~Petryk, F.~Pretorius and H.~Villegas,
  ``Towards the final fate of an unstable black string,''
  Phys.\ Rev.\ D {\bf 68}, 044001 (2003)
  [arXiv:gr-qc/0304085].

\bibitem{Horowitz}
G.~T.~Horowitz and K.~Maeda,
``Fate of the black string instability,''
Phys.\ Rev.\ Lett.\  {\bf 87}, 131301 (2001)
[arXiv:hep-th/0105111].

\bibitem{Emparan1}
R.~Emparan,
``Tubular branes in fluxbranes,''
Nucl.\ Phys.\ B {\bf 610}, 169 (2001)
[arXiv:hep-th/0105062].

\bibitem{Emparan2}
R.~Emparan and H.~S.~Reall,
``Generalized Weyl solutions,''
Phys.\ Rev.\ D {\bf 65}, 084025 (2002)
[arXiv:hep-th/0110258].

\bibitem{Emparan3}
R.~Emparan and H.~S.~Reall,
``A rotating black ring in five dimensions,''
Phys.\ Rev.\ Lett.\  {\bf 88}, 101101 (2002)
[arXiv:hep-th/0110260].

\bibitem{Emparan4}
R.~Emparan,
``Rotating circular strings, and infinite non-uniqueness of black rings,''
JHEP {\bf 0403}, 064 (2004)
[arXiv:hep-th/0402149].

\bibitem{Elvang}
H.~Elvang, R.~Emparan, D.~Mateos and H.~S.~Reall,
``Supersymmetric black rings and three-charge supertubes,''
arXiv:hep-th/0408120.

\bibitem{Dowker95}
  F.~Dowker, J.~P.~Gauntlett, G.~W.~Gibbons and G.~T.~Horowitz,
  ``The Decay of magnetic fields in Kaluza-Klein theory,''
  Phys.\ Rev.\ D {\bf 52}, 6929 (1995)
  [arXiv:hep-th/9507143].


\end{thebibliography}
\end{document}